\documentclass[11pt]{article}
\usepackage{moriond,epsfig}
\usepackage{epsfig,latexsym}
\usepackage{amsmath}
\usepackage{tensor}
\usepackage{multibox}
\usepackage{verbatim}
\usepackage{mathrsfs}
\usepackage{amssymb}
\usepackage{epsf}

\def\be{\begin{equation}}
\def\ee{\end{equation}}
\def\bea{\begin{eqnarray}}
\def\eea{\end{eqnarray}}

\begin{document}
\thispagestyle{empty}
\vspace*{1cm}

\begin{center}


{\Large\sc BRANE SUSY BREAKING AND INFLATION: \vskip 12pt IMPLICATIONS FOR SCALAR FIELDS \vskip 12pt AND CMB DISTORTION}\\


\vspace{2cm}
{\sc A.~Sagnotti${}^{\; a}$}\\[15pt]

{${}^b$\sl\small
Scuola Normale Superiore and INFN\\
Piazza dei Cavalieri, 7\\ 56126 Pisa \ ITALY \\
e-mail: {\small \it sagnotti@sns.it}}\vspace{10pt}

\vspace{2cm} {\sc\large Abstract}\end{center}
\noindent {I elaborate on a link between the string--scale breaking of supersymmetry that occurs in a class of superstring models and the onset of inflation. The link rests on spatially flat cosmologies supported by a scalar field driven by an exponential potential. If, as in String Theory, this potential is steep enough, under some assumptions that are spelled out in the text the scalar can only climb up as it emerges from an initial singularity. In the presence of another mild exponential, slow--roll inflation is thus injected during the ensuing descent and definite imprints are left in the CMB power spectrum: the quadrupole is systematically reduced and, depending on the choice of two parameters, an oscillatory behavior can also emerge for low multipoles $l < 50$, in qualitative agreement with WMAP9 and PLANCK data. The experimentally favored value of the spectral index, $n_s \approx 0.96$, points to a potentially important role for the NS fivebrane, which is \emph{unstable} in this class of models, in the Early Universe.}

\vskip 3cm
\begin{center}
{\sl Based on the talks presented at ``Rencontres de Moriond EW2013'' (La Thuile, March 2 -- 9 \ 2013), at the ``Two--Day PLANCK Meeting'', (Bologna, June 27 -- 28 \ 2013) and at the $18^{th}$ Claude Itzykson Meeting (CEA -- Saclay, July 1--3 \ 2013)}
\end{center}

\setcounter{page}{1}

\newpage

\vspace*{4cm}
\title{BRANE SUSY BREAKING AND INFLATION: IMPLICATIONS FOR SCALAR FIELDS AND CMB DISTORTION}

\author{Augusto \ SAGNOTTI}

\address{Scuola Normale Superiore and INFN \\ Piazza dei Cavalieri 7, 56126 Pisa \ ITALY\\
e-mail: \emph{sagnotti@sns.it}}

\maketitle\abstracts{I elaborate on a link between the string--scale breaking of supersymmetry that occurs in a class of superstring models and the onset of inflation. The link rests on spatially flat cosmologies supported by a scalar field driven by an exponential potential. If, as in String Theory, this potential is steep enough, under some assumptions that are spelled out in the text the scalar can only climb up as it emerges from an initial singularity. In the presence of another mild exponential, slow--roll inflation is thus injected during the ensuing descent and definite imprints are left in the CMB power spectrum: the quadrupole is systematically reduced and, depending on the choice of two parameters, an oscillatory behavior can also emerge for low multipoles $l < 50$, in qualitative agreement with WMAP9 and PLANCK data. The experimentally favored value of the spectral index, $n_s \approx 0.96$, points to a potentially important role for the NS fivebrane, which is \emph{unstable} in this class of models, in the Early Universe.}

\section{Brane SUSY breaking in String Theory}\label{sec:BSB}

Key progress in String Theory~\cite{stringtheory} in the mid nineties was spurred by the identification of dualities relating to one another spectra that appear vastly different at first sight. Some of these dualities are non perturbative from the string vantage point but find a partial justification in the low--energy Supergravity, while others are captured by string perturbation theory. The latter include the orientifold projections~\cite{orientifolds} that can associate open sectors to corresponding closed--string spectra, whose simplest manifestation is the link between the type--IIB theory of oriented closed strings and the $SO(32)$ type--I theory. In the geometrical picture proposed in~\cite{polchinski}, this particular projection is induced by spacetime--filling  non--dynamical extended objects, the $O9_-$ orientifolds, whose \emph{negative} tension $T$ and charge $Q$ are identical in suitable units. Since the corresponding lines of force would have nowhere to come from, the charge $Q$ is to be compensated via dynamical extended objects, the $D9$--branes. These carry in their turn identical tension $T$ and charge $Q$ that are however positive, so that both the total charge and the total tension cancel in the vacuum of the $SO(32)$ type--I superstring. Another option, whose significance was appreciated later, rests on a different type of orientifold, also visible in perturbation theory~\cite{witten} and whose first manifestation was found long before in~\cite{bps}. Commonly referred to as $O9_+$, this orientifold is somehow more standard, since it carries identical and \emph{positive} tension $T$ and charge $Q$. It results in a different projection \cite{sugimoto,bsb} that is still supersymmetric, but now $\overline{D}9$ anti--branes are to be present in the vacuum to compensate the positive charge, with the end result that the tensions add up rather that canceling as before while supersymmetry appears non--linearly realized in the low--energy spectrum~\cite{dmps}. More in detail, in the open sector Bose and Fermi excitations that would be paired in the SO(32) superstring display mass differences sized by the string scale $1/\sqrt{\alpha^\prime}$ and include a goldstino that conveys the breaking to the closed sector. The latter appears supersymmetric in the partition function only because in this ``brane SUSY breaking'' (BSB) phenomenon~\cite{sugimoto,bsb} the open sector emerges at a higher order in the genus expansion, from (projective) disk amplitudes, and a similar pattern is found in lower--dimensional BSB models~\cite{bsb}.

The potential applications of BSB are apparently hampered by the ``smoking gun'' that it leaves behind, an exponential potential that takes a universal form in the ``string frame'', \emph{i.e.} if the terms in the low--energy Supergravity are accompanied by powers of the string coupling
\be
g_s \ = \ e^{\,\phi} \label{string_coupling}
\ee
that reflect their origin in the Polyakov genus expansion:
\be
S_{10} = {1 \over {2\kappa_{10}^2}} \ \int d^{10} x \sqrt{-g}
\  \left\{  \, e^{\,-\,2\,\phi}\left( \, - \, R \ + \ 4 \, (\partial \phi)^2 \right)
             \ - \ T \, e^{\,-\,\phi}  \ + \ \ldots  \, \right\} \ . \label{ten_dim}
\ee
The exponential potential clearly complicates matters since flat space does not solve the field equations, and therefore insisting on the standard setting would require that resummations be implemented in String Theory~\cite{resummations}.  Still, the basic BSB phenomenon that we have illustrated has the encouraging feature of being free from tachyon instabilities at the classical level.

A vastly different option is suggested by the link introduced by BSB between the SUSY breaking and string scales, which are naturally, albeit not necessarily, identified with GUT scales ${\cal O}(10^{16})\, GeV$. Could models of this type be perhaps of interest for the Early Universe~\cite{dks,dkps}?

\section{A climbing scalar in $d$ dimensions}\label{sec:climbing}

Let us turn to consider the behavior of a minimally coupled scalar field $\Phi$ for which
\be
S \ = \ \int d^{d} x \, \sqrt{-g}\,
    \left[ \ - \ {1 \over {2\kappa_d^2}} \ R \, - \, {1 \over 2}\ (\partial \Phi)^2
            \, - \, V(\Phi) \, + \, \ldots \right]\,
\ee
in spatially flat cosmologies of the type
\be
ds^2 \, =\, - \, e^{\,
2{\cal B}(t)}\, dt^2 \, + \, e^{\, 2A(t)} \, d{\bf x} \cdot d{\bf x} \ , \qquad dt_c \ = \ e^{\,{\cal B}(t)}\, dt \ ,
\ee
where ${\cal B}(t)$ connects the  ``parametric'' time variable $t$ to the actual cosmological time $t_c$.
If the potential $V(\Phi)$ never vanishes, combining the convenient gauge choice
\be
V(\Phi) \ e^{\,2\,{\cal B}} \ = \ \frac{\overline{M}^{\ 2}}{2\, \kappa_d^{\,2}} \ \left( \frac{d-2}{d-1} \right)  \label{gauge}
\ee
with the redefinitions
\be
\tau \, = \, \overline{M} \,t\, , \ \  {\cal A} \, = \, (d-1)\, A \, , \ \
\varphi \, = \,  \kappa_d\, \sqrt{\frac{d-1}{d-2}} \, \Phi \, , \ \
\mathcal{V}(\varphi) \, = \,  2\, \kappa_d^2\, \left(\frac{d-1}{d-2}\right) \, V(\Phi) \, ,
\label{redef}
\ee
one arrives at a neat universal form for the resulting equations in an expanding Universe:
\be
\ddot{\varphi} \, + \, \dot{\varphi} \, \sqrt{1\,+\, \dot{\varphi}^{\,2}} \, +\,
\left(\, 1+ \dot{\varphi}^{\,2}\,\right)\ \frac{1}{2{\cal V}}\ \frac{\partial {\cal V}}{\partial \varphi}\, \, =\, 0 \ , \qquad
\dot{\cal A} \ = \ \sqrt{1\,+\, \dot{\varphi}^{\,2}} \ . \label{cosmo_eqs}
\ee
Here ``dots'' denote derivatives with respect to the rescaled parametric time $\tau$, and interestingly the driving force results from the \emph{logarithm} of the scalar potential.

Eqs.~\eqref{cosmo_eqs} are exactly solvable if
\be
{\cal V}(\varphi) \ = \ \left( \overline{M} \right)^{\,2} \ e^{\,2\,\gamma\, \varphi} \ ,
\ee
and many explored this type of systems after Halliwell's identification of the gauge choice \eqref{gauge}~\cite{exponential_pot}, until the exact solution was first presented for $\gamma=1$ by Dudas and Mourad in~\cite{dm} and then for all $\gamma$ by Russo in~\cite{exponential_pot}.
Let us review some key features of these solutions following~\cite{dks}, where the climbing behavior was identified, taking into account that up to redefinitions of $\varphi$ one can restrict the attention to positive values of $\gamma$. There are then two vastly different regions:
\begin{itemize}
\item[1. ] For $0 < \gamma < 1$ two distinct types of solutions exist: a \emph{climbing scalar}, for which
\be
\dot{\varphi}\, = \, \frac{1}{2}\ \left[
\sqrt{\frac{1-\gamma}{1+\gamma}}
\ \coth
\left(\frac{\tau}{2}\, \sqrt{1-\gamma^2} \right)\,-\,
\sqrt{\frac{1+\gamma}{1-\gamma}} \ \tanh
\left(\frac{\tau}{2}\, \sqrt{1-\gamma^2} \right)
\right] \ , \label{climbing_s}
\ee
and a \emph{descending scalar}, for which
\be
\dot{\varphi}\, =\, \frac{1}{2}\ \left[
\sqrt{\frac{1-\gamma}{1+\gamma}} \ \tanh
\left(\frac{\tau}{2}\, \sqrt{1-\gamma^2} \right)\,-\,
\sqrt{\frac{1+\gamma}{1-\gamma}} \ \coth
\left(\frac{\tau}{2}\, \sqrt{1-\gamma^2} \right)
\right] \ . \label{descending_s}
\ee
In the former solution $\varphi$ emerges from the initial singularity, set here at $\tau=0$, climbing up the exponential potential to then revert its motion and descend along it, while in the second it emerges directly climbing it down. In both cases, the scalar is readily driven by cosmological friction to approach the limiting speed
\be
v_l \,=\, - \
\frac{\gamma}{\sqrt{1-\gamma^{\,2}}} \ , \label{lim_speed}
\ee
and for any $0<\gamma<1$ there is also an exact solution of eq.~\eqref{cosmo_eqs} where $\varphi$ proceeds all the way at the limiting speed \eqref{lim_speed}. This is the Lucchin--Materrese (LM) attractor~\cite{lm}, which takes such a simple form in the convenient gauge \eqref{gauge}. If $\gamma < \frac{1}{\sqrt{d-1}}$ the limiting speed corresponds to a slow--roll inflationary phase of the Universe.
\item[2. ] As $\gamma \to 1$ the limiting speed diverges, while the LM attractor disappears at the ``critical'' point $\gamma=1$. The descending solution is not present anymore for $\gamma \geq 1$, where the scalar can only emerge from the initial singularity while \emph{climbing up} the corresponding steep potentials. For $\gamma=1$ the climbing solution is particularly simple, and reads
    \be
    \dot{\varphi} \, = \, \frac{1}{2\, \tau} \, -\,
\frac{\tau}{2} \ ,
    \ee
    so that for large $\tau$ it approaches a uniformly accelerated motion in the gauge \eqref{gauge}.
\end{itemize}
No additive constants are present in $\dot{\varphi}$, while $\varphi$ clearly does contain an initial--value parameter $\varphi_0$, and this can effectively \emph{tune} the strength of its interaction with the exponential barrier.

\section{String realizations}\label{sec:strin_realizations}

Can these solutions play a role in String Theory? The actual link entails an interesting subtlety, which I can briefly illustrate starting from the compactification of the Lagrangian \eqref{ten_dim} to $d$ dimensions on the metric
\be
ds^2 \, = \, e^{\,-\, \frac{(10-d)}{(d-2)} \,\sigma} \ g_{\mu\nu}\, dx^\mu\, dx^\nu \ + \ e^{\,\sigma} \ \delta_{ij}\, dx^i \, dx^j \ ,
\ee
whose dependence on the scalar $\sigma$ that sizes the internal volume has been arranged in such a way that the system ends up in the Einstein frame. The reduced Lagrangian,
\be
S_{d} \, = \, {1 \over {2\kappa_{d}^2}} \,  \int d^{\,d} x \sqrt{-g}
\  \left\{  \, - \, R \, - \, \frac{1}{2} \ (\partial \phi)^2
\, - \, \frac{2(10-d)}{(d-2)} \ (\partial \sigma)^2
 \, - \, T \ e^{\,\frac{3}{2} \,\phi\, -\,  \frac{(10-d)}{(d-2)}\,
\sigma} \, + \, \ldots  \right\}
\ee
can be turned into the more conventional form
\be
S_{d} \ = \  {1 \over {2\kappa_{d}^2}} \ \int d^{\,d} x \sqrt{-g}
\  \left\{  \ - \  R  \  -  \
\frac{1}{2} \, (\partial \Phi_s)^2 \  -  \
\frac{1}{2} \, (\partial \Phi_t)^2 \ - \ T \ e^{\,\Delta \, \Phi_t}\ + \ \ldots  \right\} \label{s_d}
\ee
by field redefinitions but then, remarkably, rescalings similar to those in eq.~\eqref{redef} show that the exponential potential for $\Phi_t$ has $\gamma=1$, and is thus ``critical'' for all $d$~\cite{fss}! The presence of the second scalar $\Phi_s$ clearly complicates matters, but we shall \emph{assume} nonetheless that it is somehow stabilized and we shall thus follow the common practice of concentrating on one--field models of inflationary Cosmology.

A climbing scalar is of special significance in String Theory, since it is naturally compatible with an upper bound on the dilaton $\phi$ and thus with a perturbative string regime. However, while later epochs will be central in what I am about to describe, let me stress that I am not aware of fully convincing arguments to ignore, as we did in \cite{dks,dkps}, higher--derivative corrections to the effective action \eqref{ten_dim} near the initial singularity, which generally make climbing not inevitable. Nonetheless, let me conclude this section on a positive note, mentioning briefly another little miracle~\cite{dks}: in four dimensions the climbing behavior persists even if one includes the axion partner $\theta_t$ of $\Phi_t$, since its non--minimal kinetic term freezes it out near the initial singularity.

\begin{figure}
\begin{center}$
\begin{array}{cc}
\epsfig{file=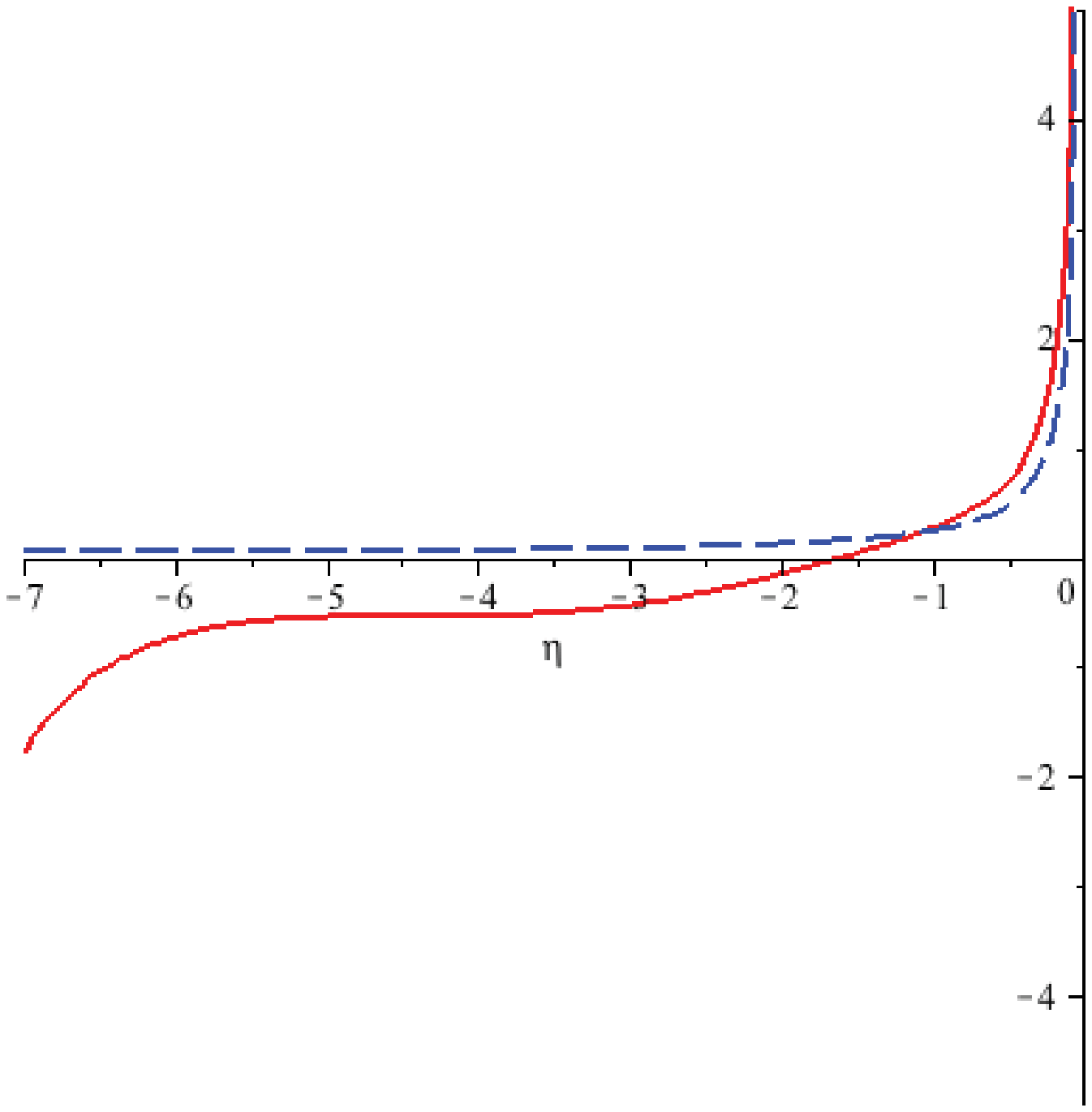, height=1.5in}\qquad \qquad \qquad \qquad &
\epsfig{file=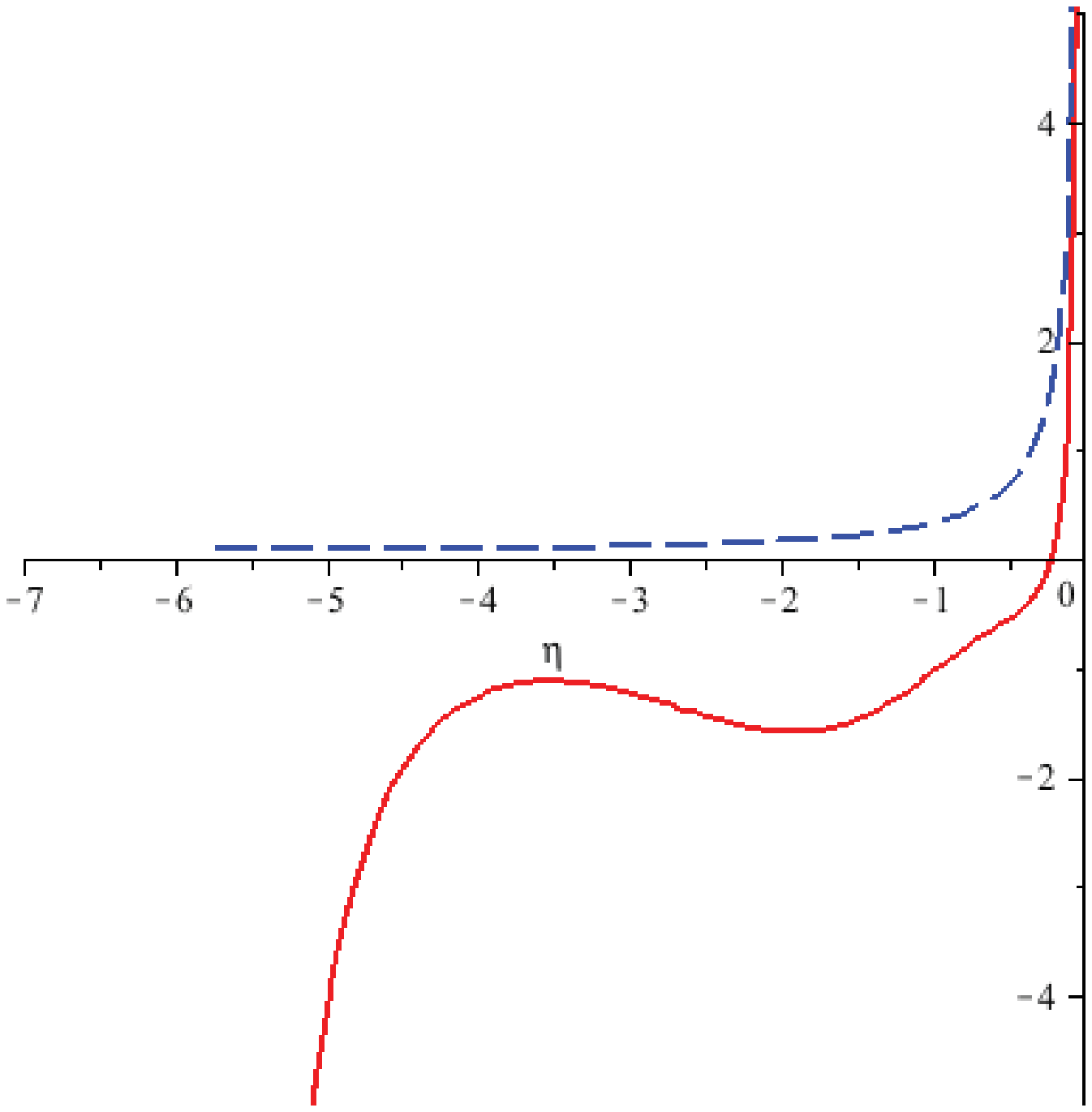, height=1.5in}
\end{array}$
\end{center}
\caption{MS potentials for the two--exponential case of eq.~\eqref{two_exp_pot} with $\varphi_0=-4$ (left) and with $\varphi_0=0$ (right). Both approach eventually the LM attractor curve (dashed line). Notice, however, that this occurs earlier in the first case, where the curve also overtakes it, and later in the second, where the curve always stays well below.
\label{fig:comparisonWs}}
\end{figure}

\section{Implications for the CMB power spectrum}\label{sec:CMB_power}

The critical exponential potential of eq.~\eqref{s_d} is not alone in String Theory. Already in the simple model of~\cite{sugimoto} it is accompanied in principle by a similar term with $\gamma=1/2$ that originates from the non--BPS D3 brane of~\cite{dms} and is capable of supporting an inflationary phase, so that in the following I shall focus on the more general class of potentials
\be
{\cal V}(\varphi) \, = \, \overline{M}^{\ 2}  \bigg( e^{\, 2\, \varphi} \ + \ e^{\, 2 \, \gamma\, \varphi} \bigg) \ , \label{two_exp_pot}
\ee
and the comparison with the actual CMB power spectrum tilt determines $\gamma \approx \frac{1}{12.4}$ as an optimal choice. This class of potentials combines an early climbing phase, a sort of bounce against a ``hard exponential wall'' and a final inflationary descent. It is not exactly solvable in general, but two choices of qualitatively similar integrable potentials are described in~\cite{fss}. The wide scan presented in~\cite{br} opens a number of possibilities for brane contributions to be held responsible for the value needed to account for the tilt, and an extension of the reasoning sketched in the preceding section yields a prediction for the values of $\gamma$ that can be induced by a generic $p$ brane coupling to the dilaton, in string frame, as $\exp(\,-\,\alpha\,\phi)$. The result is simply \cite{fss}
\be
\gamma \ = \ \frac{1}{12} \ \left( p \ + \ 9 \ - \ 6\, \alpha \right) \ , \label{gammabranes}
\ee
so that these values are remarkably quantized in units of $\frac{1}{12}$, a few percents from the experimentally favored value! There is also a clear suspect for the best--fit value $\frac{1}{12}$, the NS fivebrane wrapped on a small internal cycle ($p=4$, $\alpha=2$). This brane is interestingly \emph{unstable} in orientifold models, so that it is tempting to associate to its decay the graceful exit from slow roll and the subsequent reheating of the Universe.

Let me now turn to examine the implications of the potential \eqref{two_exp_pot} for the CMB scalar power spectrum. The key tool is provided by the Mukhanov--Sasaki (MS) equation~\cite{cosmo_reviews},
\be
\frac{d^{\,2} v_{k}(\eta)}{d\eta^2} \ + \ \big[ k^2\ - \ W_s(\eta) \big] v_{k}(\eta) \, = \, 0 \ ,
\ee
where
\be ds^2 \, =\, e^{\,\frac{2}{3}\,{\cal A}(\eta)} \, \left( \, -\, d \eta^2 \, +\, d{\bf x} \cdot d{\bf x} \right) \ , \qquad W_s = \frac{1}{z} \ \frac{d^{\,2} \, z}{d\eta^{\,2}} \ , \qquad z(\eta) \, \sim \, e^{\,\frac{1}{3} \,{\cal A}(\eta)} \, \frac{d\,\varphi(\eta)}{d\,{\cal A}(\eta)}  \ . \ee
$\varphi(\eta)$ and ${\cal A}(\eta)$ are background values and $\eta$ denotes the conformal time.
Details on the spectrum of tensor perturbations, which also overshoots the attractor curve and disappears as $k\to 0$, can be found in~\cite{dkps}.
\begin{figure}
\begin{center}$
\begin{array}{cc}
\epsfig{file=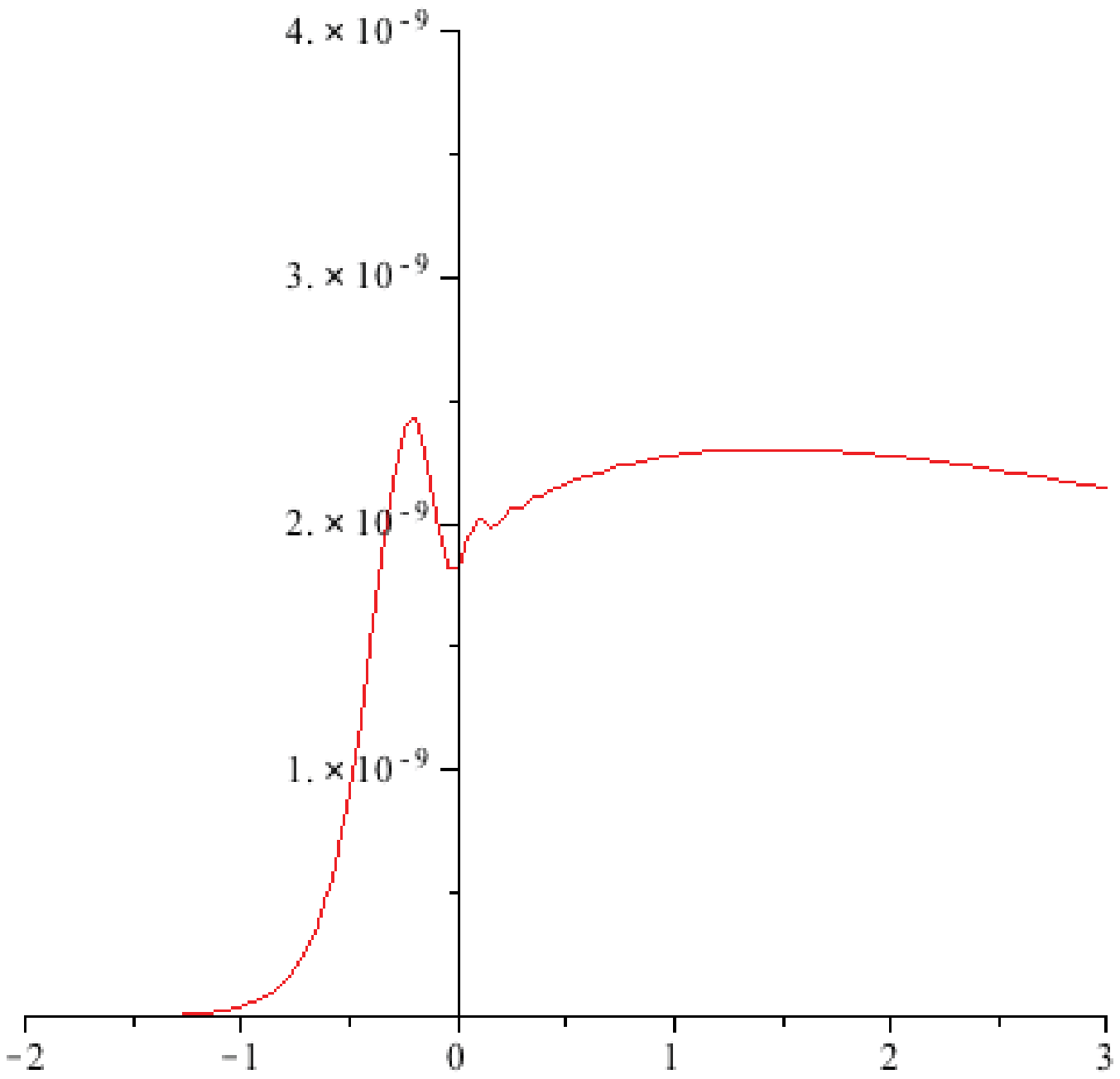, height=1.5in}\qquad \qquad &
\epsfig{file=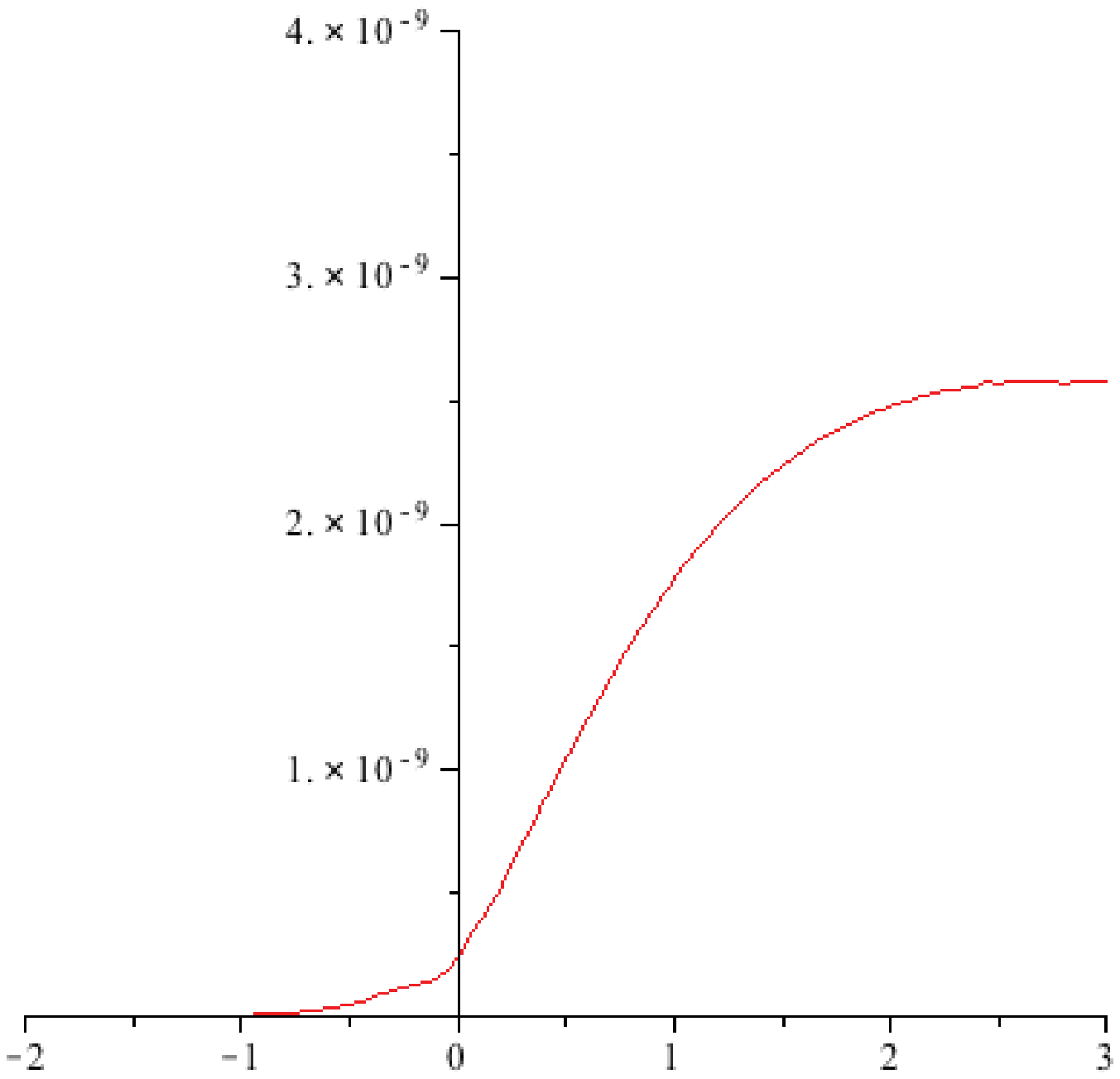, height=1.5in}
\end{array}$
\end{center}
\caption{Scalar power spectra for two values of $\varphi_0$, $\varphi_0=-1.5$ (left) and  $\varphi_0=0$ (right).
\label{fig:spectrum}}
\end{figure}

The evolution described by the MS equation finds an instructive analogy in the time--independent \emph{boundary--value} Schr\"odinger problem, with the important proviso that in inflationary dynamics one is actually solving an \emph{initial--value} problem for the counterparts of the flat--space exponentials $e^{-iE_k t}$. The MS potential $W_s(\eta)$ has some universal features, since it behaves near the initial singularity (at a finite negative conformal time $-\eta_0$) and at late times ($\eta \to 0^-$) as
\be
W_s \ \ {\phantom a}_{\widetilde{\eta \to - \eta_0}} \ \ - \ \frac{1}{4} \, \frac{1}{(\eta+\eta_0)^2} \ , \qquad W_s \ \ {\phantom a}_{\widetilde{\eta \to 0^-}} \  \ \ \frac{\nu^2\,-\, \frac{1}{4}}{\eta^2}  \qquad
\left[  \nu \ =  \ \frac{3}{2} \ \frac{1 \, - \, \gamma^{\, 2}}{1 \ - \ 3\, \gamma^{\, 2}}  \right] \ . \label{W_s}
\ee
As a result, $W_s$ must cross the real axis, and actually does it once in the models of interests, before approaching an infinite barrier at the origin of conformal time (fig.~1). In Quantum Mechanics this barrier would result in total reflection, but in the MS \emph{initial--value} problem the growing mode generally dominates in the classically forbidden region. In other words, the WKB ``barrier penetration factor'' leaves way here to a ``barrier amplification factor'', and after a (large) number of $e$--folds that
depends on $\epsilon$
\be
v_{k}(\,-\,\epsilon)\ \sim \ \frac{1}{\sqrt[4]{ |W_s(\,-\,\epsilon)\,-\, k^2|}} \ \exp\left(\int_{-\eta^\star}^{-\epsilon} \sqrt{|W_s(y)\,-\, k^2| }\, dy \right) \ ,
\ee
where $-\eta^\star$ denotes the classical inversion point. The extent of the amplification reflects the area below the positive portion of $W_s$, and therefore an inspection of fig.~\ref{fig:comparisonWs} suffices to acquire a clear qualitative picture of the power spectrum
\be
P(k) \ \sim  \ k^3 \ \left| \frac{v_k(\,-\,\epsilon)}{z(\,-\,\epsilon)} \right|^2 \ .
\ee

The plots in fig.~\ref{fig:comparisonWs} show typical MS potentials $W_s$ for the two--exponential problem and for a potential ${\cal V}(\varphi)$ containing only the milder term, and finally in all cases the dashed curves correspond to the LM attractor, for which the second of eqs.~\eqref{W_s} applies for all negative $\eta$'s.
Notice that $P(k)$ must tend to zero as $k\to 0$ simply due to the initial singularity, which forces the curve to cross the real axis, so that the area below it is bounded as $k \to 0$. As a result, the power spectra for our ``climbing'' systems experience a $k^3$--falloff for small $k$, in contrast with the $k^{3-2\nu}$--growth that occurs for the LM attractor. On the other hand, for large $k$ the ``climbing'' power spectra approach the attractor result, albeit more slowly in the two--exponential system, whose $W_s$ stay well below the attractor curves of fig.~\ref{fig:comparisonWs} for a while. These considerations are well reflected in fig.~\ref{fig:spectrum}, but for the oscillations that are missed altogether by the WKB approximation, here as in Quantum Mechanics, where they would reflect resonant transmission through a potential well.

\begin{figure}
\begin{center}$
\begin{array}{cc}
\epsfig{file=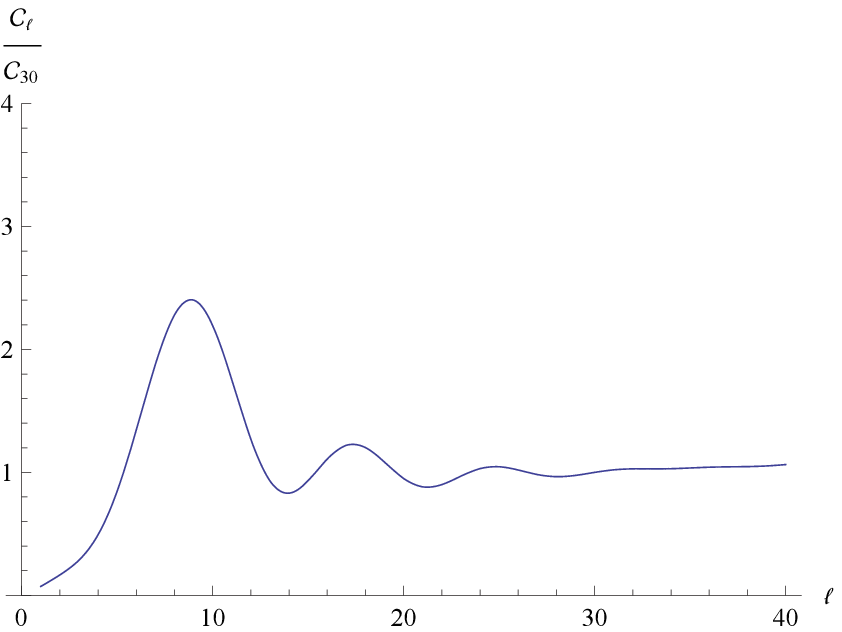, height=1.5in}\qquad \qquad &
\epsfig{file=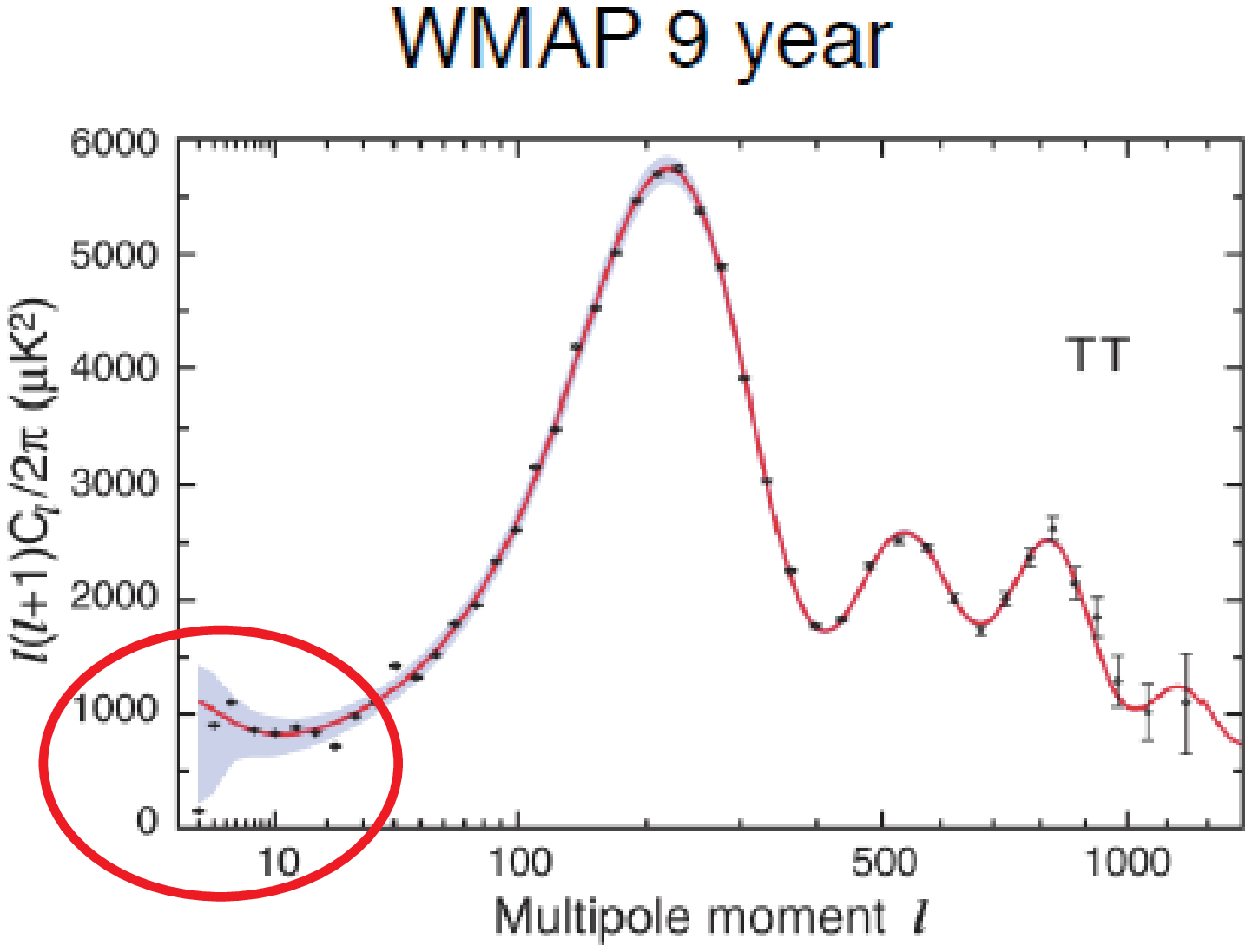, height=1.5in}
\end{array}$
\end{center}
\caption{A qualitative comparison between the low--$\ell$ portion of the WMAP9 plot and the first $C_\ell$'s for the BSB--inspired potential \eqref{two_exp_pot}, normalized with respect to $C_{30}$ and computed for a climbing phase that occurred about one $e$--fold before the horizon exit of the current Hubble scale. The oscillations are very sensitive to $\varphi_0$ and disappear if the current Hubble scale exited more than 3--4 $e$--folds after the onset of inflation.
\label{fig:comparison}}
\end{figure}

\section{An observable window in the Cosmic Microwave Background ?}

Can this class of string--inspired models capture some features of the WMAP9 or PLANCK multipole plots~\cite{wmap9}? The actual comparison depends, so to speak, on the portion of the power spectra of~\cite{dkps} that is accessible to current observations. One can anticipate that any significant effects should only concern the low--$\ell$ portions, since the power spectra of fig.~\ref{fig:spectrum} merge eventually with the attractor curve, but our real chance of connecting the current data to String Theory via BSB rests on the enticing possibility that Nature is unveiling the onset of inflation. The low--$k$ portions of the power spectra translate directly, via the Fourier--Bessel integrals
\be
C_\ell \ = \ \frac{2}{9\pi} \ \int \frac{dk}{k} \ P(k) \ j_\ell^2\big[k\, \Delta \eta\big] \ ,
\ee
where $\Delta \eta$ denotes our current comoving distance from the last scattering surface,
into corresponding predictions for the multipole coefficients with $\ell<50$. Since the squared $j_\ell$'s are peaked for arguments of order $\ell$, if our Universe were confronting us with the growing portions in fig.~\ref{fig:spectrum} one could anticipate that the quadrupole should be reduced for all models under scrutiny. On the other hand, the behavior of subsequent multipoles should depend on the details of the dynamics, and thus on the value of $\varphi_0$. In \cite{dkps} we contented ourselves with the quadrupole reduction, but playing with $\varphi_0$ can enhance the oscillations, so that one can end up with curves like the left one of fig.~\ref{fig:comparison}. This is qualitatively similar to the low--$\ell$ portion of the WMAP9 results, which is surrounded by the ellipse in fig.~\ref{fig:comparison}, so that String Theory and BSB are perhaps finding some indirect evidence in the CMB! Of course, cosmic variance adds more than a word of caution to this suggestion, but nonetheless one can explore the possibility of arriving at a best fit of the present data playing with the two parameters at our disposal, the observable window of the spectrum and the value of $\varphi_0$. The optimal model would be an ideal starting point to analyze the bispectrum, which could then lend further support to this picture (or perhaps disprove it)~\cite{dkps2}. Let me conclude by stressing that refined analyses of the low--$\ell$ tail of the CMB power spectrum are starting to appear \cite{gruppuso}, and that they point to a lowering of the quadrupole. Time will tell whether these exciting signs will materialize.

\section{Conclusion}

\begin{figure}
\begin{center}$
\begin{array}{cc}
\epsfig{file=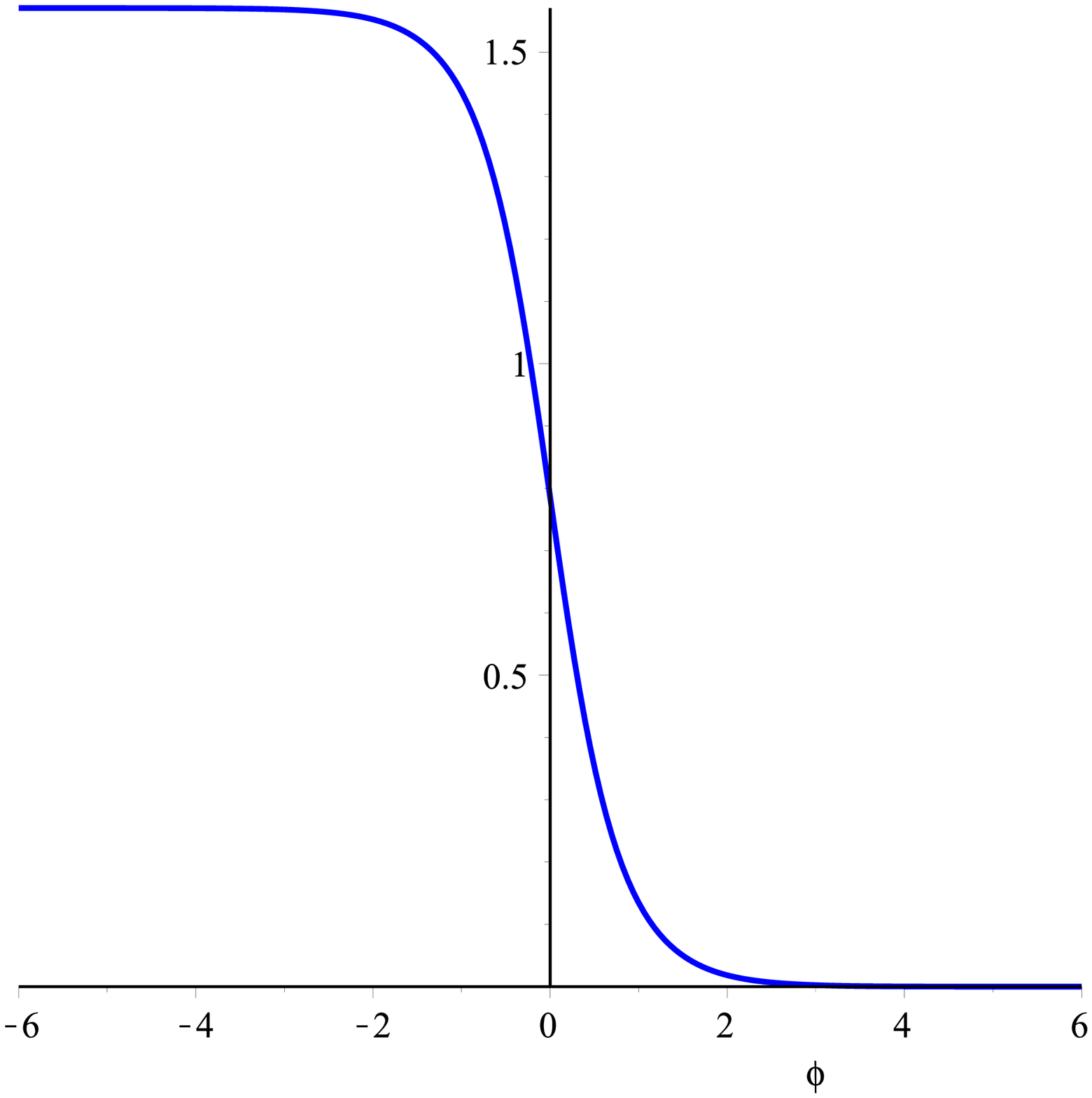, height=1.2in, width=1.2in}\qquad \qquad &
\epsfig{file=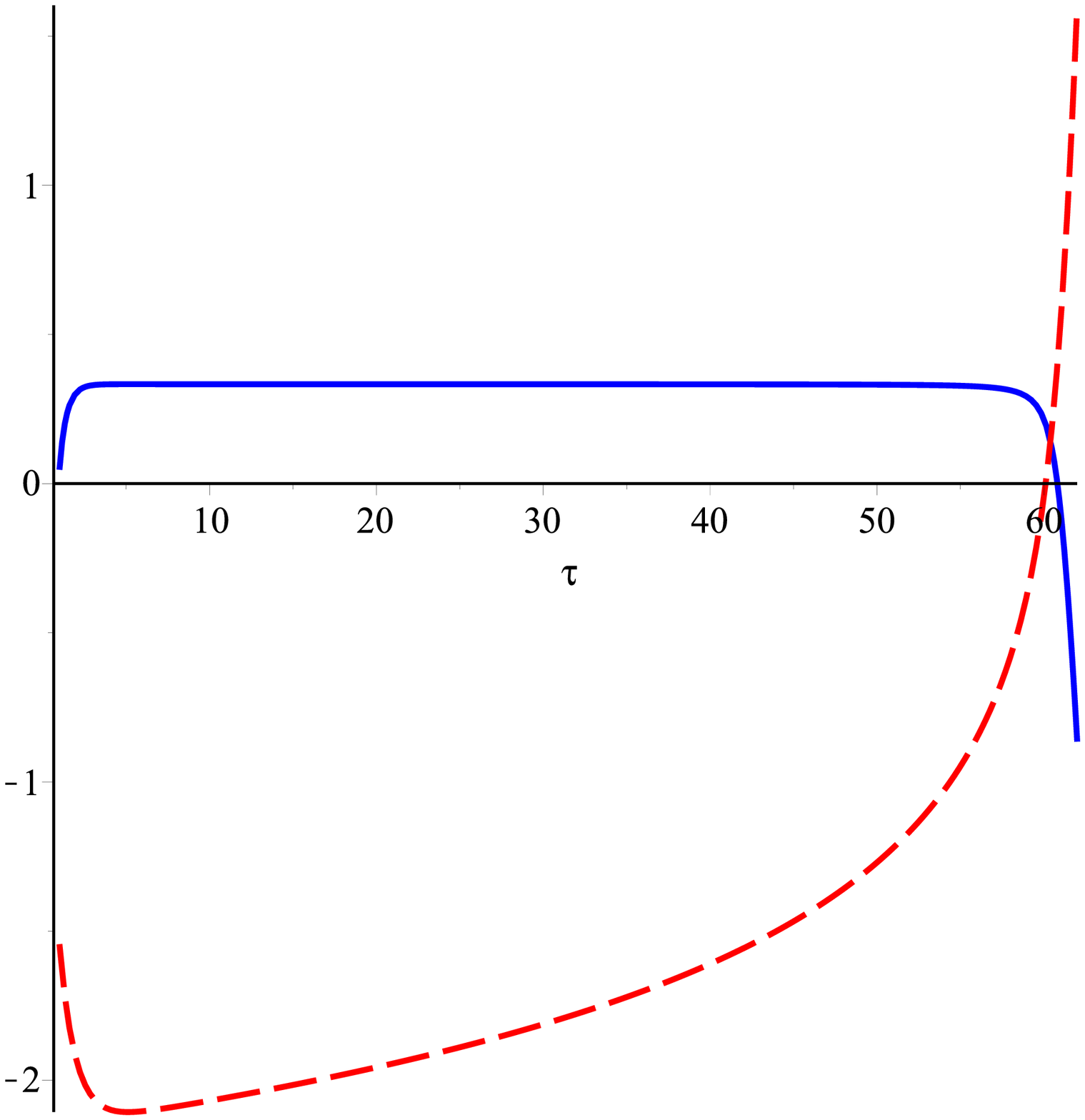, height=1.2in, width=1.2in}
\end{array}$
\end{center}
\caption{The step potential of eq.~\eqref{step_pot} (left) and a graceful exit from climbing and inflation (right): the dotted curve and the continuous one represent, respectively, $\varphi(t)$ and the acceleration of the Universe.
\label{fig:step}}
\end{figure}

I have reviewed the work of~\cite{dks}, where a link was proposed between a peculiar string--scale SUSY breaking mechanism, ``brane SUSY breaking'' or BSB for short, and the onset of inflation. I have also reviewed its application to the CMB power spectrum presented in~\cite{dkps}, and I have mentioned some recent results that are in qualitative agreement with the low--$\ell$ tails of WMAP9 or PLANCK data. BSB results in a ``critical'' logarithmic slope for a tree--level exponential potential, and under some assumptions this forces the inflaton (a mixture of the dilaton and the scalar related to the volume of the extra dimensions, in the setting that we have analyzed) to emerge from the initial singularity while climbing it up. The subsequent descent could have injected the inflationary phase of our Universe, so that \emph{String Theory and BSB are perhaps providing some clues onwhy and how inflation started}. Remarkably, under the same assumptions all branes in String Theory yield tree--level contributions to the scalar potential with \emph{logarithmic slopes that are quantized in terms of} $\gamma = \frac{1}{12}\,$, which lies a few percents away from the experimentally favored $\frac{1}{12.4}$! As we have seen, this picture could have left tangible signs in the CMB power spectrum that are intriguingly along the lines of the WMAP9 plot of fig.~\ref{fig:comparison}. How about the subsequent evolution, then? Admittedly, we are not addressing in detail key issues like the graceful exit and reheating, since our current grasp of String Theory would be of little help in this respect, although eq.~\eqref{gammabranes} points to the (unstable) NS fivebrane, which could have played a key role in connection with the graceful exit from inflation and with the subsequent reheating. At any rate, the relevant scalar actors of the early phase couple to other fields in the rich fashion that is typical of Supergravity, in a version with non--linear supersymmetry but containing nonetheless the types of matter couplings that are generally associated with reheating (see \emph{e.g.}~\cite{fk} and references therein). More work is needed to clarify the issue, but let me close mentioning a remarkable exact solution~\cite{fss} whose potential (left portion of fig.~\ref{fig:step})
\be
{\cal V}(\varphi) \ \sim \ \arctan\left(e^{\,-\,2\,\varphi} \right) \label{step_pot}
\ee
combines a ``critical'' tree--level exponential with similar, if ad hoc, higher--genus closed--string terms to provide a vivid picture of a graceful exit from an initial climbing phase. This potential is essentially a step function with a slight tilt and $\varphi$ has the option of emerging from the right to climb it up, linger for quite a while on the plateau and then eventually roll down as inflation ends. Or, alternatively, to emerge from the left, undergo slow roll on the plateau and roll down as inflation ends. The right portion of fig.~\ref{fig:step} displays an example where 50 $e$--folds of inflation are produced climbing up from the right. The early climbing phase, however, is not inevitable in this example: the scalar could also move fast all the way, giving rise to no inflation at all.
\section*{Acknowledgments}
I am grateful to E.~Dudas, P.~Fr\'e, S.P.~Patil, F.~Riccioni, A.S.~Sorin and especially to N.~Kitazawa (who also produced some improved power spectra displayed in this upgraded version of the talk) for several stimulating discussions, and to CERN, the \'Ecole Polytechnique and the Lebedev Institute for their kind hospitality. I am also grateful to the Organizers of Rencontres de Moriond EW2013 for the very stimulating Conference, for giving me the opportunity to present these results and for letting me update the file, at the cost of exceeding slightly their page limit, so that the paper now contains the main results also presented at the ``Two--Day PLANCK Meeting'' held in Bologna on June 27-28 2013. Finally, I would like to thank E.~Calabrese and A.~Melchiorri, who kindly sent me a PLANCK multipole plot, and A.~Gruppuso and P.~Natoli for their interest in this work. This research was supported in part by the ERC Advanced Grants n. 226455 (SUPERFIELDS) and n. 226371 (MassTeV), by Scuola Normale Superiore, by INFN (I.S. TV12) and by the MIUR-PRIN contract 2009-KHZKRX.


\section*{References}
\begin{thebibliography}{99}
\bibitem{stringtheory} For a recent review see: K.~Becker, M.~Becker and J.~H.~Schwarz,
  ``String theory and M-theory: A modern introduction,''
  Cambridge, UK: Cambridge Univ. Pr. (2007) 739 p.
\bibitem{orientifolds}
A.~Sagnotti, in Cargese '87, ``Non-Perturbative Quantum Field
Theory'', eds. G. Mack et al (Pergamon Press, 1988), p. 521,
arXiv:hep-th/0208020;
G.~Pradisi and A.~Sagnotti,
Phys.\ Lett.\ B {\bf 216} (1989) 59;
P.~Horava,
Nucl.\ Phys.\ B {\bf 327} (1989) 461,
Phys.\ Lett.\ B {\bf 231} (1989) 251;
M.~Bianchi and A.~Sagnotti,
Phys.\ Lett.\ B {\bf 247} (1990) 517,
Nucl.\ Phys.\ B {\bf 361} (1991) 519;
M.~Bianchi, G.~Pradisi and A.~Sagnotti,
Nucl.\ Phys.\ B {\bf 376} (1992) 365;
A.~Sagnotti,
 Phys.\ Lett.\  B {\bf 294}, 196 (1992)
 [arXiv:hep-th/9210127]~.
For reviews see: E.~Dudas,
Class.\ Quant.\ Grav.\  {\bf 17}, (2000) R41 [arXiv:hep-ph/0006190];
C.~Angelantonj and A.~Sagnotti,
Phys.\ Rept.\  {\bf 371} (2002) 1 [Erratum-ibid.\  {\bf 376} (2003)
339] [arXiv:hep-th/0204089].

\bibitem{polchinski}
 J.~Polchinski,
  Phys.\ Rev.\ Lett.\  {\bf 75} (1995) 4724
  [hep-th/9510017].

\bibitem{witten}
 E.~Witten,
  JHEP {\bf 9802} (1998) 006
  [hep-th/9712028].

\bibitem{bps}
M.~Bianchi, G.~Pradisi and A.~Sagnotti, in \cite{orientifolds}.

\bibitem{sugimoto}
S.~Sugimoto,
Prog.\ Theor.\ Phys.\  {\bf 102} (1999) 685 [arXiv:hep-th/9905159].

\bibitem{bsb}
I.~Antoniadis, E.~Dudas and A.~Sagnotti,
Phys.\ Lett.\ B {\bf 464} (1999) 38 [arXiv:hep-th/9908023];
C.~Angelantonj,
Nucl.\ Phys.\ B {\bf 566} (2000) 126 [arXiv:hep-th/9908064];
G.~Aldazabal and A.~M.~Uranga,
JHEP {\bf 9910} (1999) 024 [arXiv:hep-th/9908072];
C.~Angelantonj, I.~Antoniadis, G.~D'Appollonio, E.~Dudas and
A.~Sagnotti,
Nucl.\ Phys.\ B {\bf 572} (2000) 36 [arXiv:hep-th/9911081].

\bibitem{dmps}
 E.~Dudas and J.~Mourad,
  Phys.\ Lett.\ B {\bf 514} (2001) 173
  [hep-th/0012071];
G.~Pradisi and F.~Riccioni,
  Nucl.\ Phys.\ B {\bf 615} (2001) 33
  [hep-th/0107090].

\bibitem{resummations}
W.~Fischler and L.~Susskind,
 Phys.\ Lett.\  B {\bf 171}, 383 (1986),
 Phys.\ Lett.\  B {\bf 173}, 262 (1986);
E.~Dudas, M.~Nicolosi, G.~Pradisi and A.~Sagnotti,
 Nucl.\ Phys.\  B {\bf 708} (2005) 3
 [arXiv:hep-th/0410101];
N.~Kitazawa,
 Phys.\ Lett.\  B {\bf 660} (2008) 415
 [arXiv:0801.1702 [hep-th]].

\bibitem{dks}
E.~Dudas, N.~Kitazawa and A.~Sagnotti,
  Phys.\ Lett.\ B {\bf 694} (2010) 80
  [arXiv:1009.0874 [hep-th]].


\bibitem{dkps}
  E.~Dudas, N.~Kitazawa, S.P.~Patil and A.~Sagnotti,
  JCAP {\bf 1205} (2012) 012
  [arXiv:1202.6630 [hep-th]].


\bibitem{exponential_pot}
J.J.~Halliwell,
  Phys.\ Lett.\  {\bf B185} (1987) 341;
L.F.~Abbott and M.B.~Wise,
 Nucl.\ Phys.\  {\bf B244} (1984) 541;
D.H.~Lyth and E.D.~Stewart,
 Phys.\ Lett.\  {\bf B274} (1992) 168;
 E.~Dudas and J.~Mourad,
 Phys.\ Lett.\  {\bf B486} (2000) 172
 [arXiv:hep-th/0004165];
I.P.C.~Heard and D.~Wands,
  Class.\ Quant.\ Grav.\  {\bf 19} (2002) 5435
  [arXiv:gr-qc/0206085];
N.~Ohta,
  Phys.\ Rev.\ Lett.\  {\bf 91} (2003) 061303
  [arXiv:hep-th/0303238];
S.~Roy,
  Phys.\ Lett.\  {\bf B567} (2003) 322
  [arXiv:hep-th/0304084];
P.K.~Townsend and M.N.R.~Wohlfarth,
  Phys.\ Rev.\ Lett.\  {\bf 91} (2003) 061302
  [arXiv:hep-th/0303097],
 Class.\ Quant.\ Grav.\  {\bf 21} (2004) 5375
  [arXiv:hep-th/0404241];
R.~Emparan and J.~Garriga,
  JHEP {\bf 0305} (2003) 028
  [arXiv:hep-th/0304124];
J.G.~Russo,
  Phys.\ Lett.\  {\bf B600} (2004) 185
  [arXiv:hep-th/0403010];
A.A.~Andrianov, F.~Cannata and A.Y.~Kamenshchik,
  JCAP {\bf 1110} (2011) 004
  [arXiv:1105.4515 [gr-qc]],
  arXiv:1206.2828 [gr-qc].

\bibitem{dm}
 E.~Dudas and J.~Mourad,
 Phys.\ Lett.\  B {\bf 486} (2000) 172
 [arXiv:hep-th/0004165].

\bibitem{lm}
F.~Lucchin and S.~Matarrese,
 Phys.\ Rev.\  {\bf B32} (1985) 1316.

\bibitem{fss} P.~Fr\'e, A.~Sagnotti and A.~S.~Sorin,
  arXiv:1307.1910 [hep-th].

  \bibitem{dms}
 E.~Dudas, J.~Mourad and A.~Sagnotti,
 Nucl.\ Phys.\  B {\bf 620} (2002) 109
 [arXiv:hep-th/0107081].

\bibitem{br}
E.~A.~Bergshoeff and F.~Riccioni,
  JHEP {\bf 1105} (2011) 131
  [arXiv:1102.0934 [hep-th]],
  arXiv:1109.1725 [hep-th];
E.~A.~Bergshoeff, A.~Marrani and F.~Riccioni,
  Nucl.\ Phys.\ B {\bf 861} (2012) 104
  [arXiv:1201.5819 [hep-th]].

\bibitem{cosmo_reviews}
  V.~Mukhanov, ``Physical foundations of cosmology,''
  Cambridge, UK: Univ. Pr. (2005) 421 p;
  S.~Weinberg, ``Cosmology,''
  Oxford, UK: Oxford Univ. Pr. (2008) 593 p;
  D.~H.~Lyth and A.~R.~Liddle,
  ``The primordial density perturbation: Cosmology, inflation and the origin of structure,''
  Cambridge, UK: Cambridge Univ. Pr. (2009) 497 p.

\bibitem{wmap9}
G.~Hinshaw, D.~Larson, E.~Komatsu, D.~N.~Spergel, C.~L.~Bennett, J.~Dunkley, M.~R.~Nolta and M.~Halpern {\it et al.},
  arXiv:1212.5226 [astro-ph.CO];
    C.~L.~Bennett, D.~Larson, J.~L.~Weiland, N.~Jarosik, G.~Hinshaw, N.~Odegard, K.~M.~Smith and R.~S.~Hill {\it et al.},
  arXiv:1212.5225 [astro-ph.CO];
  P.~A.~R.~Ade {\it et al.}  [ Planck Collaboration],
  arXiv:1303.5062 [astro-ph.CO].

\bibitem{dkps2}
E.~Dudas, N.~Kitazawa, S.P.~Patil and A.~Sagnotti, in progress.

\bibitem{gruppuso}
A.~Gruppuso, P.~Natoli, F.~Paci, F.~Finelli, D.~Molinari, A.~De Rosa and N.~Mandolesi,
  arXiv:1304.5493 [astro-ph.CO].

\bibitem{fk}
S.~Ferrara and R.~Kallosh,
  JHEP {\bf 1112} (2011) 096
  [arXiv:1110.4048 [hep-th]].


\end{thebibliography}
\end{document}